\documentclass{pasj00}

\begin{document}
\SetRunningHead{Deguchi et al.}{Circumstellar Chemistry of the Bipolar Nebula IRAS 19312+1950}
\Received{2004/07/01}
\Accepted{2004/09/29}

\title{Study of the Bipolar Nebula IRAS 19312+1950. II. 
Circumstellar Chemistry}

\author{Shuji \textsc{Deguchi}$^{1,2}$,  Jun-ichi \textsc{Nakashima}$^{2, 3}$}
\and
\author{Shuro \textsc{Takano}$^{1,2}$}

\affil{$^1$ Nobeyama Radio Observatory, National Astronomical Observatory \\
Minamimaki, Minamisaku, Nagano 384-1305}
\affil{$^2$ Department of Astronomical Science, Graduate University for Advanced Studies,  \\
Nobeyama Radio Observatory, Minamimaki, Minamisaku, Nagano 384-1305}
\affil{$^3$ Department of Astronomy, University of Illinois at Urbana/Champaign, \\ 
1002 W Green St., Urbana, Illinois 61801, U. S. A.}



%


\maketitle

\begin{abstract}
The bipolar nebula, IRAS 19312+1950,  is a unique  SiO maser source 
exhibiting both properties as young and evolved objects. 
To clarify the nature of this object,  we made molecular line observations 
with the Nobeyama 45-m radio telescope. We detected emission from O-bearing
(HCO$^{+}$,  SiO,  SO, and SO$_2$),    
C- and N-bearing molecules  (CN, CS, HCN, HNC, NH$_{3}$, N$_{2}$H$^{+}$, 
HC$_{3}$N, H$_2$CS, and CH$_3$OH), and their isotopic species
(C$^{17}$O,  $^{13}$C$^{18}$O,  and C$^{34}$S).  
Line profiles consist of  a weak broad 
($\Delta v\sim 30$ km s$^{-1}$) and/or  a strong narrow ($\Delta v\lesssim 5$ km s$^{-1}$)
components, depending on molecular species.
Strong time variations of H$_2$O and SiO masers 
were also observed.  Numerical modeling of the envelope 
with the LVG-code results in a good fit of the model with a mass loss
rate of $2.6 \times10^{-4} M_{\odot}$ yr$^{-1}$  to
the observed intensities for the broad-component lines.  On the other hand, 
non-O-bearing molecules, which have only the narrow profiles,
were found to have abundances typical to  those in  cool dust clouds. 
No isotopic enrichment was found, indicating little evidence of the narrow cool 
component being an ejecta of  the central AGB star or a possible companion.  
These facts compelled us to conclude  that 
IRAS 19312+1950 is an exotic mass-losing evolved star embedded in 
a low-mass ($\sim 20 \; M_{\odot}$)  dark cloud. 
\end{abstract}

\section{Introduction}
IRAS 19312+1950  exhibits a prominent bipolar nebulosity 
of the size of about 30$''$ in near-infrared images  (\cite{nak00}). 
The IRAS color of this source, $log(F_{25}/F_{12})=0.47$, indicates a presence of cool dust 
of a temperature of about 200 K toward this source.
Detecting SiO and H$_{2}$O maser lines in this object, \citet{nak00} concluded that this is
an oxygen-rich protoplanetary nebula similar to OH231.8+4.2.  
The distance to this object was estimated to be about 2.5 kpc  (\cite{nak04}; hereafter Paper I),
based on the near- and mid-IR flux densities. 
It is slightly unusual that CO and the other thermal molecular lines of less abundant species are
detectable in the envelope of an Asymptotic Giant Branch (AGB) or post-AGB stars at
 2.5 kpc away from the sun. For example, 
the distance to the similar object, OH231.8+4.2, is about 1.3 kpc 
(\cite{mor87}), and those of  other commonly observed AGB stars in radio 
are less than 1 kpc (\cite{olo98}). 
Near-IR imaging of this object with the UH 88-inch telescope 
shows a bright central star and surrounding nebula, extending 
from NE to SW by about 30$''$ (figure 1) . 
It also shows a small ($\sim 5''$) ring-like
structure near the central star, which is elongated perpendicularly 
to the bipolar axis. 

In paper I,  we reported the results of mapping observations by several molecular lines
(CO, HCN, and CS) using the Nobeyama 45-m radio telescope; the line
profiles exhibit broad and/or narrow components depending on molecular species.
We found  that both broad and narrow components
centered at $V_{\rm lsr}\simeq$ 35 km s$^{-1}$
are strongly peaked in intensity at the IRAS position of this source
with a 16$''$ telescope beam.
Without doubt, the broad component with a full width of more than 20 km s$^{-1}$ 
must originate from the outflowing envelope of the central star.
However, the spatial association of the narrow component to this object
is a puzzle. The mass of the narrow component was estimated to be
about 10 $M_{\odot}$ (within 16$''$ beam). The mass obtained is compatible with 
the core masses of dark clouds (for example, \cite{ful91}).

In this paper, we present a part of  molecular-line observations 
toward this object, and discuss on it from  a view point of circumstellar chemistry.
In usual cases, molecular species found in dark clouds are considerably different from
those found in the evolved objects [see, \citet{gla96} and \citet{ehr00}]. 
Therefore, the origin of the
cloud surrounding the central star of IRAS 19312+1950 can be resolved
from this study in some degree.  
Although evidence of an evolved star is still not completely crucial for this object 
or rather conflicting, 
we believe that this is a unique evolved object
surrounded by a remnant of raw material. We present the observational results and
discussion on the nature of this object in this paper.

\setcounter{figure}{0}
\begin{figure}
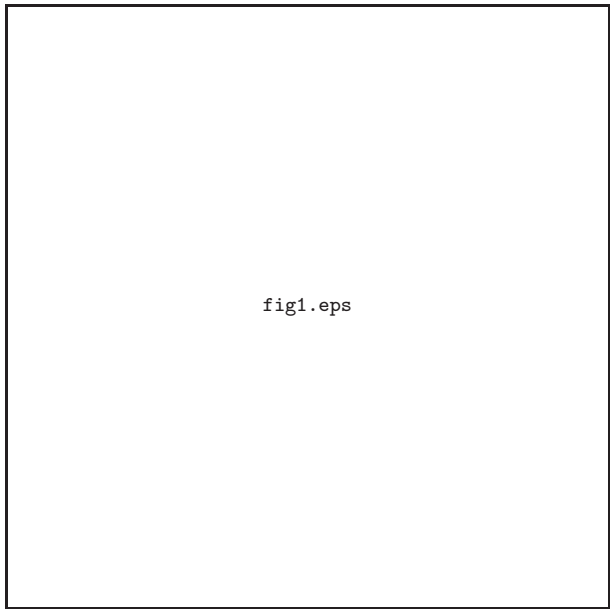

  \begin{center}
    \FigureFile(80mm,80mm){fig1.eps}
  \end{center}
  \caption{$JHK$ false-color composite image of the field of
  IRAS 19312+1950, using the UH 88-inch telescope 
  with the SIRIUS near-IR array camera
  (by courtesy of M. Tamura and K. Murakawa, National Astronomical Observatory). 
  The image size is $128''\times 128''$.
  The north is up, and the east is left.}\label{fig1}
\end{figure}


\section{Observations and results}

The observations were made with the 45-m telescope at Nobeyama
during 2001 February 4--9, and during 2002 December 11--21.
A few additional observations were made on 2001 March 19 and 2002 May 19
for checking time variation of SiO maser line intensities. 
A cooled SIS-mixer receiver (S100) with a bandwidth of about 0.5 GHz were used, 
and the system temperature (including atmospheric noise) was 300--500 K (SSB)
depending on the frequency and weather.  In addition,
HEMT and SIS-mixer receivers (H22 and S40)
were used for  the 22 GHz H$_2$O and the 43/49 GHz SiO/CS 
line observations, respectively. 
The half-power beam widths, and the aperture and beam efficiencies of the telescope 
at the observed frequencies
were summarized in table 1. The antenna temperature ($T_{a}^*$) 
given in the present paper is that corrected for the atmospheric 
and telescope ohmic loss but not for the beam nor aperture efficiency.
Observations  were made in a position switching mode,  
and the off-position was chosen 12$'$ away from the
object in right ascension, where contamination by 
background CO emission was minimized. 
Further details of the observations were described in Paper I.

{ \setlength{\tabcolsep}{2pt}\footnotesize
\begin{table}
  \caption{Telescope parameters (2001 March)}\label{tab:table1}
 \begin{center}
  \begin{tabular}{lrrcc}
  \hline\hline
Receiver&Frequency&HPBW&Beam &Aperture\\
        &         &    &efficiency&efficiency\\
&(GHz)&($''$)&&\\
	\hline
H22&22&72&0.81&0.62\\
S40&43&40&0.77&0.57\\
S100&86&18.6&0.49&0.42\\
S100&100&16.3&0.51&0.41\\
S100&110&15.4&0.48&0.35\\
   \hline
  \end{tabular}
  \end{center}
\end{table} }

As described in Paper I,  strong CO emission was seen toward this object.
The 15$''$-,  30$''$-, and 60$''$-grid mappings revealed
that emission spikes at $V_{\rm lsr}=27$--33, and 43 km s$^{-1}$, 
are extended more than a few arcmin, indicating that
they are from fore/background clouds (see Paper I).
We found that only the component at $V_{\rm lsr}=$37 km s$^{-1}$
is sharply peaked in intensity at the IRAS position  
($19^{\rm h}31^{\rm m}12^{\rm s}$.8, 19$^{\circ}$50$'$ 22$''$, B1950).
Mapping of line emissions of high density tracers,
 HCN, HCO$^{+}$, and CS, resulted in
that the component at $V_{\rm lsr}=$35--37 km s$^{-1}$
is sharply peaked also at the IRAS position of this source.
In this paper, we only show a result of
the 5-point 1$'$-grid cross mapping in the HCO$^{+}$ $J=1$--0 transition
in figure 2. We can recognize from this figure 
that the H$^{12}$CO$^{+}$ $J=1$--0
line consist of the two components: a narrow component
at $V_{\rm lsr}=$37 km s$^{-1}$, and the weak broad component
spreading from $V_{\rm lsr}=$18 km s$^{-1}$ to 50 km s$^{-1}$. 
Because these two components are also found 
in the other molecular lines as CO, HCN, and SO, 
and because they are strongly concentrated on the IRAS position
of this object, we conclude that they are
not from the fore/background clouds.
Because strong lines as the CO $J=$1--0, $^{13}$CO $J=$1--0, 
C$^{18}$O $J=1$--0, CS $J=2$--1, and HCN $J=1$--0 (Paper I), are somewhat
contaminated by emission from the back/foreground 
molecular clouds,  we presented the result of the mapping by these lines separately
in Paper I.  The results for HCO$^+$ described in this paper, i.e.,  the narrow and broad components and 
their spatial extensions, were completely confirmed by mm-wave interferometric observations using 
the BIMA array (\cite{nak04b})

\begin{figure}
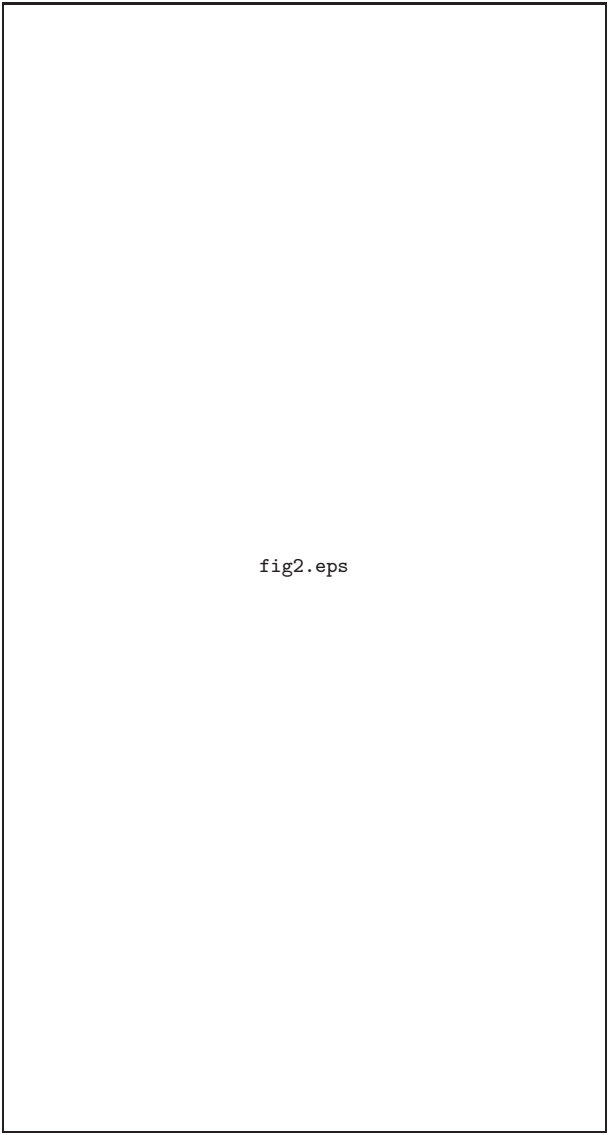

  \begin{center}
    \FigureFile(80mm,150mm){fig2.eps}
  \end{center}
  \caption{Line profiles of HCO$^{+}$ 
  and H$^{13}$CO$^{+}$ $J=1$--0 transitions. The observed position
   for each spectrum is indicated in the parentheses 
 [ ($\Delta$R.A., $\Delta$Dec.) relative to the IRAS position] on the right.
 The HCO$^{+}$ profile at the center clearly shows broad 
 ($V_{\rm lsr}=20$--50 km s$^{-1}$) and
 narrow (around $V_{\rm lsr}=36$ km s$^{-1}$) components.}\label{fig2}
\end{figure}

\subsection{Time variation of maser lines}

We  also observed  maser lines
to check their time variation. The results are summarized in table 2
and the line profiles are shown in figure 3. 
It was striking that the SiO $J=1$--0 $v=2$ spectrum in 2001 February--March  
(middle two panels in figure 3) exhibited
strong emission at $V_{\rm lsr}\simeq 50$ km s$^{-1}$, but
almost none in the $J=1$--0 $v=1$ and $J=2$--1 $v=1$ spectra.
The integrated intensity of the SiO $J=2$--1 $v=1$ line 
increased by a factor of more than 3 in about 40 days in 2001 March.
If we compare these line profiles with those taken in May 2000 
[figure 2 in \citet{nak00}],
we can recognize drastic changes in the line profiles during 8 months.
In May 2000, the radial velocities of most of maser lines
were crowded around $V_{\rm lsr}\simeq 20$ km s$^{-1}$. 
However, we observed in 2001 February 
that they spread between $V_{\rm lsr}\simeq 20$ km s$^{-1}$
and 55 km s$^{-1}$. This suggests that the center velocity of the star
is $\sim 37$ km s$^{-1}$, when we assume  these maser lines 
are formed in the front and rear parts of the expanding shell.

\begin{figure*}
  \begin{center}
    \FigureFile(110mm,130mm){fig3.eps}
  \end{center}
  \caption{Time variation of the H$_{2}$O and SiO maser lines. Dates of observations
are shown  in yymmdd format under the object name in each panel.}\label{fig3}
\end{figure*}

 Top two panels of figure 3  show a drastic time variation 
 in H$_2$O maser profiles during one-year time span.
 In 2001 February, emission appeared around the stellar velocity, $\sim 37$ km s$^{-1}$. 
 While, after one year later,
 the profile looked doubly peaked and the line intensity became larger . 
 Such a rapid change of the H$_2$O maser line profile between the 
groups A (single peak at stellar velocity) and group B (double peaks) in water masers 
(\cite{eng86})  indicates that the direction of maser amplification changes in a short time
(\cite{deg77}).  If we assume that the double peaks
of H$_2$O masers  are emitted in the front and rear part of
the expanding envelope,  we can derive $V_{\rm lsr}=$33.6 km s$^{-1}$
for the systemic velocity  and  17.3 km s$^{-1}$ for the expansion velocity
of the H$_2$O maser shell.   The intensity of the lower velocity component tends 
to surpass that of the higher velocity component in circumstellar H$_2$O masers,
indicating a maser amplification of continuum radiation of the central star (\cite{tak01}).


Emissions of OH 1612, 1665, 1667 MHz lines have been detected
toward this object (Lewis 2000, private communication).
Though the profiles of these OH lines are not exactly doubly peaked, 
emissions spread between $V_{\rm lsr}=$10 and 55 km s$^{-1}$. This 
observation also suggests that the stellar velocity 
is around $V_{\rm lsr}\simeq 33$ km s$^{-1}$, which agrees with
the stellar velocity derived from SiO and H$_2$O profiles within uncertainty.

{\setlength{\tabcolsep}{2pt}\footnotesize
\begin{table*}
  \caption{Time variations of maser line intensities.}\label{tab:table2}
  \begin{center}
  \begin{tabular}{lccrrrrr}
  \hline\hline
Mol.&Frequency&Transition&$V_{\rm lsr}$(Peak)&$T_{a}^*$(Peak)&$Integ. Inten.$ &rms&Obs.date\\
& (GHz)&&(km s$^{-1}$)&(K)&K (km s$^{-1}$)&(K)&(yymmdd)\\
	\hline
H$_2{}$O&22.235 &6(16)--5(23)&35.6&2.588&12.105&0.031&010210\\
 &&&&&&&\\ 
H$_2{}$O&22.235&6(16)--5(23)&16.3&6.628&12.526&0.323&021216\\
H$_2{}$O&22.235&6(16)--5(23)&50.8&0.372& 0.607&0.323&021216\\
&&&&&&&\\
SiO&43.122&1--0 $v=$1&41.1&0.108&0.122&0.021&010209\\
&42.821&1--0 $v=$2&49.9&0.184&0.817&0.021&010209\\
&86.243&2--1 $v=$1&36.9&0.184&0.513&0.020&010209\\
&&&&&&&\\
SiO&43.122&1--0 $v=$1&---&---&---&0.033&010319\\
&42.821&1--0 $v=$2&53.8&0.328&1.347&0.042&010319\\
&86.243&2--1 $v=$1&23.2&0.199&1.731&0.020&010319\\
&&&&&&&\\
SiO&43.122&1--0 $v=$1&51.8&0.213&0.376&0.055&020519\\
     &42.821&1--0 $v=$2&51.0&0.414&1.036&0.067&020519\\
&&&&&&&\\
SiO&43.122&1--0 $v=$1&50.8&0.186&0.346&0.030&021215\\
     &42.821&1--0 $v=$2&51.7&0.420&0.970&0.043&021215\\
   \hline
  \end{tabular}
  \end{center}
\end{table*}  }

\subsection{Thermal lines of oxygen-bearing molecules}
From the detections of SiO, H$_{2}$O and OH maser lines,
we can safely suggest that the circumstellar envelope of this object 
is  O-rich. However, the detections of H$^{13}$CN $J=1$--0 line
toward this object and the velocity shift to the maser lines
(\cite{nak00}) puzzled us, 
and these facts yielded the argument if the thermal lines
may be due to contamination by background.

To clarify the situation, we observed several
oxygen-bearing molecules as SiO ($J=1$--0, $v=0$),
SO, SO$_{2}$, and HCO$^{+}$. The results are summarized in table 3,
and the line profiles are shown in figure 4.
In addition, several lines of the various O-bearing molecules 
were searched for;
frequencies of some transitions of these molecules were
fallen in our spectrometer band coverages.
We carefully checked whether or not any
emission was seen in the spectra. 
These negative results are summarized in table 4.

The line profiles of these O-bearing molecules (except H$^{13}$CO$^{+}$) 
are composed of two components: narrow sharp feature
peaked at $V_{\rm lsr}\simeq 36$ km s$^{-1}$, and the broad feature
with a full width at zero maximum of about 30 km s$^{-1}$ (see figures 2 and 4).
It seems that the SiO $J=2$--1 $v=0$ line profile does not have any narrow
component. For the case of the SO$_2$ 10(1,9)--10(0,10) line, the detection 
is marginal and it is difficult to tell whether it has the narrow or broad component.

{\setlength{\tabcolsep}{2pt}\footnotesize
\begin{longtable}{lrcllllllll}
  \caption{Line parameters for line with broad and narrow features}\label{tab3}
  \hline\hline
 & & &\multicolumn{3}{c}{peak feature}&\multicolumn{3}{c}{broad compo.}
     &\multicolumn{2}{c}{narrow compo.}\\
   \cline{4-6} \cline{7-9} \cline{10-11}
Mol.&Freq.&Transition&$V_{\rm lsr}({\rm p})$&$T^*_a({\rm p})$&rms&$V_{\rm lsr}({\rm par})$&$T^*_a({\rm par})$&FWZM&$T^*_a(nar)$&I.I.(nar)\\
 &{\tiny(GHz)}& &{\tiny (km s$^{-1}$)}&{\tiny (K)}&{\tiny(K)}&{\tiny (km s$^{-1}$)}&{\tiny(K)}&{\tiny (km s$^{-1}$)}&{\tiny (K)}&{\tiny (K km s$^{-1}$)}\\
\endfirsthead
	\hline
SiO&86.847&2--0 v=0&33.3&0.128&0.027&35.1&0.086&34.9&---&---\\
H$^{13}$CO$^{+}$&86.754&1--0&36.2&0.123&0.021&---&---&---&0.123&0.556\\
HCO$^{+}$&89.189&1--0&36.9&0.326&0.025&32.3&0.060&41.8&0.269&0.849\\
SO&99.300&3(2)--2(1)&35.8&0.624&0.019&33.5&0.112&44.5&0.513&2.459\\
  &109.252&2(3)--1(2)&35.6&0.104&0.043&36.2&0.028&35.2&0.076&0.093\\
  &100.030&4(5)--4(4)&---&---&0.023&---&---&---&---&---\\
SO$_2$&104.239&10(1,9)--10(0,10)&35.8&0.040&0.012&---&---&---&---&---\\
	\hline
\end{longtable} }

It is well known that the profile of the
optically thin line in the expanding shell
is flat-topped and that of the thick line is parabolic
when the telescope beam cannot resolve the envelope (\cite{mor77}).
The profiles of the broad components in figure 4 
do not seem to be flat-topped, but rather they look
like parabolic. Therefore, 
we fitted the profiles of the broad component
by a parabolic line-shape function 
(excluding the narrow part of emission)
and obtained the line parameters; the results
are given in the 7--9 columns in table 3.
The line parameters for the narrow components,
which were computed from the residue by the parabolic fit, 
are also given in the last two columns in table 3.

The average center velocity of the broad components,
34.3 km s$^{-1}$, is very slightly shifted from the average
peak velocity of the narrow component, 35.6 km s$^{-1}$.
In addition the profile of the SiO thermal line ($J=2$--1 $v=$0)
in figure 4 looks asymmetric with respect to the center velocity.
The difference in the radial velocity
between the narrow and board components seems significant.
However, because the broad component is quite weak and 
overlapped with the narrow component, it is difficult to separate
the broad component feature from the narrow component clearly
in the present observations. Moreover, the profile of the CO $J=1$--0 line
(top right panel of figure 1 of Paper I) clearly indicates a presence of 
absorptions on both high and low velocity sides of the narrow component,
but with a slightly stronger absorption in the lower velocity side.
Therefore, the  difference of the center-velocity between the narrow and broad 
components could be due to radiation transfer effect (absorption at lower velocity side) 
by a weakly expanding part of the cloud with low excitation temperature (for example, \cite{mor85}).

\begin{figure}
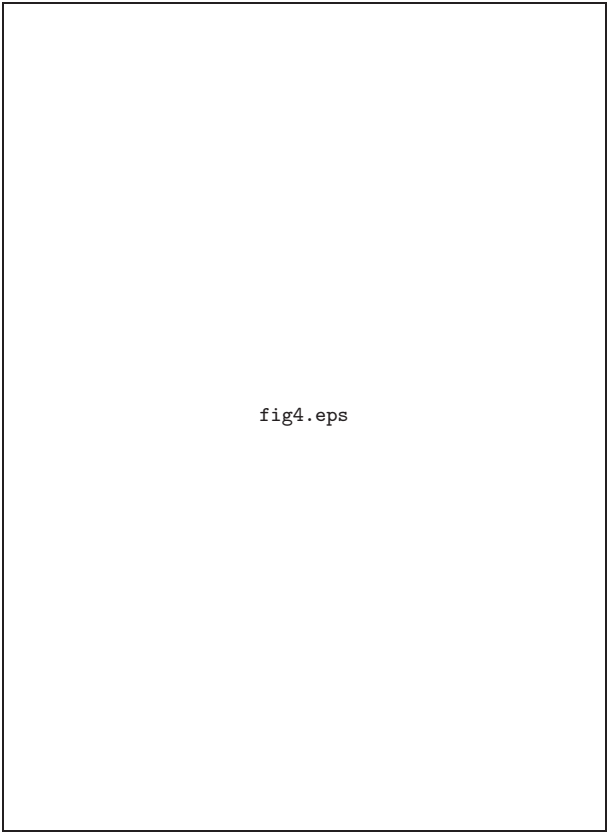

  \begin{center}
    \FigureFile(80mm,110mm){fig4.eps}
  \end{center}
  \caption{Line profiles of SiO $J=2$--1 $v=$0,
  SO 3(2)--2(1) and 2(3)--1(2), and SO$_2$ 10(1,9)--10(0,10) transitions.
  The SiO profile  exhibits only a broad component, and
  SO profiles exhibit both broad and narrow components. 
  Because of the noise.  it is hard to judge whether it is a narrow or 
  broad component for SO$_2$.
 }\label{fig4}
\end{figure}

\begin{table}
  \caption{Negative results for S-bearing molecules}\label{tab4}
  \begin{center}
   \begin{tabular}{lrcl}
  \hline\hline
Mol.&Freq.&Transition&rms\\
&(GHz)&&(K)\\
	\hline
SO$_{2}$&86.153&39(9,31)--40(8,32)&0.016\\
&86.639&8(3,5)--9(2,8)&0.025\\
&90.548&25(3,23)--24(4,20)&0.049\\
&104.029&3(1,3)--2(0,2)&0.020\\
&&&\\
$^{34}$SO$_{2}$&88.720&7(3,5)--8(2,9)&0.042\\
&104.391&10(1,9)--10(0,10)&0.025\\
&&&\\
$^{34}$SO&97.715&3(2)--2(1)&0.063\\
&&&\\
OCS&109.463&9--8&0.023\\
&&&\\
SiS&90.771&5--4&0.058\\
&&&\\
C$^{33}$S     & 48.585   &     1--0      &   0.033 \\
&&&\\
C$^{34}$S & 48.206    &   1--0       &  0.027  \\
\hline
  \end{tabular}
  \end{center}
\end{table}

\subsection{Thermal lines of C- and N-bearing molecules}
We have also detected a number of lines of 
C- and N-bearing molecules. The results
are summarized in table 5, and line profiles are
shown in figures 5 and 6. These lines often have
hyperfine splitting except the CS lines.
The spectra of these lines do not seem to have
the broad component but they exhibit the narrow
component, although the hyperfine splitting
(shown by arrows in figure 6) tends to conceal the
presence of the broad component.

Line widths  of narrow components at the half intensity maxima 
are approximately 1.5--2.0 km s$^{-1}$. The average 
center velocity of these lines in table 5 is 35.9 km s$^{-1}$,
which is consistent with the average velocity of 
the narrow component of O-bearing molecules.
The width of the narrow component
seems to vary from molecule to molecule.
The CS line has a full width of about 8 km s$^{-1}$,
while, HCO$^{+}$ less than 2 km s$^{-1}$.
This fact can be interpreted by a characteristic 
of the lines formed in a turbulent media where the
width depends on the line optical depth [for example, see \citet{par95}].

{\setlength{\tabcolsep}{2pt}\footnotesize
\begin{longtable}{lrccccc}
  \caption{Line parameters for species with narrow features only}\label{tab5}
  \hline\hline
Mol.&Freq.&Transition&$V_{\rm lsr}(\rm{peak})$&$T^*_a(\rm{peak})$&Integ.Int.&rms\\
& (GHz)&&(km s$^{-1}$)&(K)&(K km s$^{-1}$)&(K)\\
\endfirsthead
	\hline
NH$_{3}$&23.694&(1,1)&35.1&0.091&0.731&0.012\\
&23.723&(2,2)&37.4&0.041&0.203&0.010\\
&23.870&(3,3)&---&---&---&0.010\\
&24.139&(4,4)&---&---&---&0.022\\
&&&&&&\\
CS&48.991&1--0&36.2&0.621&1.977&0.042\\
&97.981&2--1&36.3&1.100&3.852&0.059\\
&&&&&&\\
C$^{34}$S&96.413& 2--1& 35.8&0.191&0.441&0.0118\\
&&&&&&\\
H$_2$CS&104.617&3(1,2)--(2(1,1)&35.1&0.046&0.313&0.010\\
&&&&&&\\
HNC&90.600&1--0&35.4&0.581&1.930&0.039\\
&&&&&&\\
N$_{2}$H$^{+}$&93.174&1--0&34.1&0.122&0.657&0.020\\
&&&&&&\\
HC$_{3}$N&100.076&11--10&36.3&0.163&0.481&0.018\\
HC$_{3}$N&109.174&12--11&36.2&0.104&0.403&0.020\\
&&&&&&\\
CN&113.491&1-0 (all)&36.7&0.521&3.439&0.033\\
&&&&&&\\
CH$_3$OH&96.739 & 2($-1$)--1(1) E & 35.4 & 0.142& 0.384 & 0.018 \\
CH$_3$OH&96.741 & 2(0)--1(0) A+ & 35.9 & 0.159& 0.448 & 0.018 \\
&&&&&&\\
C$^{17}$O&112.359&1--0 (+hyperfine)&35.5&0.197&1.101&0.027\\
$^{13}$C$^{18}$O&104.711&1--0 & 36.4 &0.045&0.200&0.010\\
	\hline
\end{longtable} }

The molecular species exhibiting the narrow component are somewhat 
similar to those observed in the C-rich circumstellar
envelopes (e.g., \cite{buj94}). A number of carbon-chain molecules
have been detected in the C-rich protoplanetary nebulae
(\cite{fuk94}). However, SiS, which is common in
the C-rich envelopes, was not detected in IRAS 19312+1950.

The high HNC/HCN abundance ratio is often considered
as a signature of photochemical reactions
(\cite{mor87}; \cite{gla96}). The line intensity of the HNC
emission, $T_{a}^*\sim 0.6$ K, is 
comparable with the HCN intensity, $T_{a}^*\sim 0.6$ K
in this object, suggesting a photochemical origin
of this molecule. Even taking into account 
the line saturation in the HCN lines
[$\tau (F=2-1)\sim $2--4 can be deduced from 
$T_{a}^*(F=0-1)/T_{a}^*(F=2-1)=0.56$ and 
$T_{a}^*(F=1-1)/T_{a}^*(F=2-1)=0.76$ in this star, where
these ratios are 0.2 and 0.6,
respectively, in a thermal equilibrium], 
we can reach the same conclusion that the HNC/HCN abundance ratio 
is relatively high ($>0.3$) in the narrow component. 

The molecular species  exhibiting the narrow component are also 
similar to those found in the dark clouds; 
NH$_3$, SO, H$_2$CS , and CH$_3$OH,  etc., were observed in  LDN 134N [see \citet{dic00}].
Therefore, in order to check the similarity in molecular abundances to dark clouds, we also searched for   H$_2$CS and CH$_3$OH.
We found emission from these molecules and the line profiles are shown
in  the bottom panels of  figures 5 and  6.

{\setlength{\tabcolsep}{2pt}\footnotesize
\begin{longtable}{lrlll}
  \caption{Negative results for miscellaneous species}\label{tab6}
  \hline\hline
 Molecule & Frequency  &   Transition  &  rms &   Freq. Ref. \\
 &(GHz)&&(K)&\\
\endfirsthead
	\hline
 MgNC & 23.859   &     2--1$- $  &   0.006  &   \cite{kaw93} \\
            & 23.875    &    2--1+ &     0.006 &   \cite{kaw93}  \\
            & 47.726   &     4--3$- $  &   0.023  &    \cite{kaw93} \\
            & 47.741    &    4--3+   &   0.023  &   \cite{kaw93} \\
		    &  95.454     &     $N=8$---7$- $  &  0.016  &   \cite{gue95} \\
            &  95.469       &    $N=8$---7+   &  0.016  &    \cite{gue95} \\
 &&&&\\
$^{26}$AlCl   &  89.351   &    6--5    &  0.016  & calc.   from \citet{gue95} \\
              & 104.237    &   7--6    &  0.013  &   calc.   from \citet{gue95}  \\ 
&&&&\\
 CH$_3$OH  & 48.372  &   1(0)--0(0) A+  & 0.028  &  \cite{lov92} \\
               & 48.376  &   1(0)--0(0) A+  & 0.028  &  Lovas 1992 \\
&&&&\\
C$_4$H & 104.667 & 11--10 J=23/2--21/2 & 0.010 &   \cite{lov92} \\
               & 104.705 & 11--10 J=21/2--19/2 & 0.010 &   \cite{lov92} \\
 &&&&\\
H$_2$O    & 96.261    & 4(4,0)--5(3,3) $\nu_2=1$  & 0.018 &   \cite{lov92} \\
\hline
\end{longtable} }

\begin{figure}
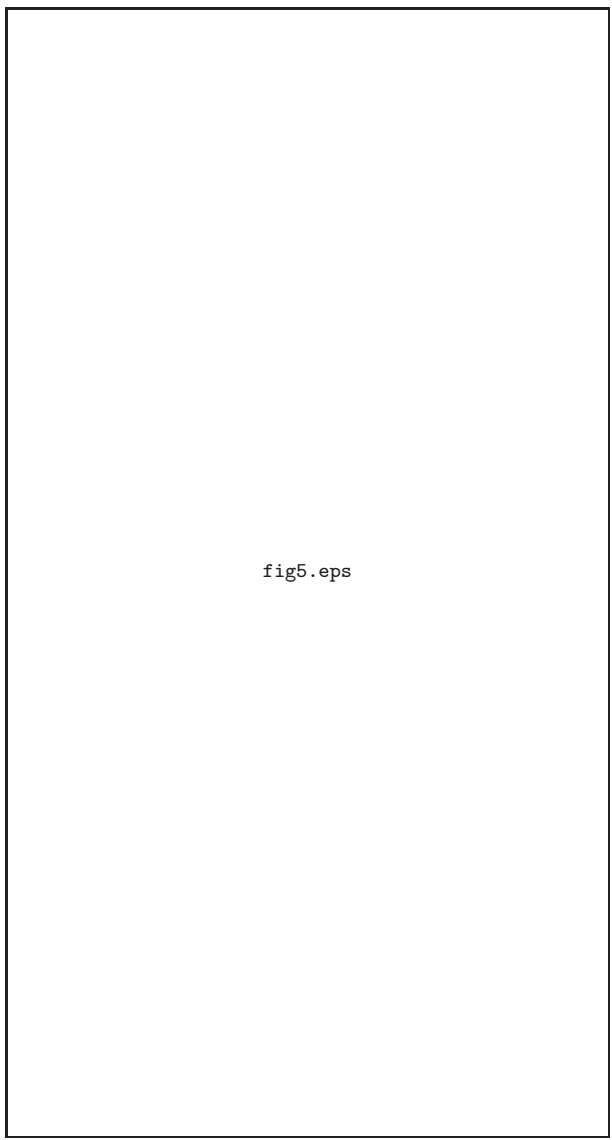

  \begin{center}
    \FigureFile(80mm,150mm){fig5.eps}
  \end{center}
  \caption{Line profiles of HC$_{3}$N, CS and H$_2$CS. The hyperfine
  splitting of HC$_{3}$N is quite small and not recognizable here.
  Both of HC$_{3}$N and CS lines seem to exhibit a narrow component.}\label{fig5}
\end{figure}
\begin{figure}
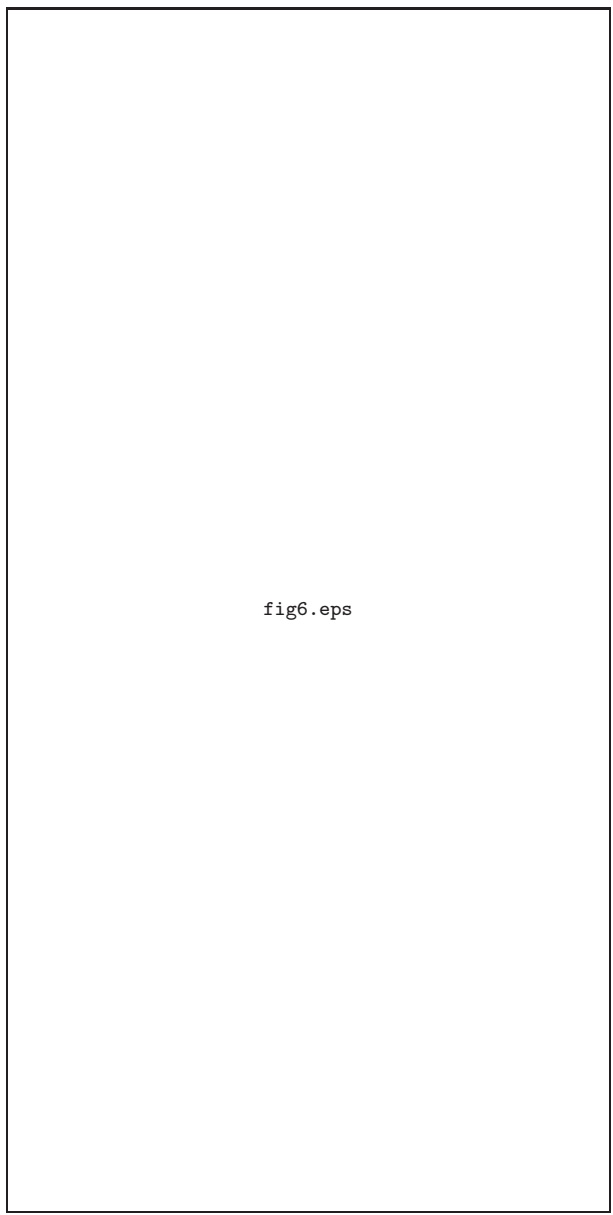

  \begin{center}
    \FigureFile(80mm,160mm){fig6.eps}
  \end{center}
  \caption{Line profiles of NH$_{3}$, HNC, N$_{2}$H$^{+}$, and CH$_3$OH.
  The positions of the hyperfine lines were indicated by arrows.
  The splitting of HNC are less than 0.7 km s$^{-1}$, 
  so that they are not shown here.
  All of these lines exhibits only a narrow components.}\label{fig6}
\end{figure}

\section{Discussion}

Paper I  estimated the distance to this object as 2.5 kpc;  the bolometric
correction for the IRAS 12 micron flux density,  
$(BC)_{12}=$3.8 (\citet{deg01}),  was applied from the $R$ (DSS), 
 $J$, $H$, and $K$ (2MASS) magnitudes and IRAS 12, 25,
60, and 100 $\mu$m flux densities, and 
typical luminosity of the AGB object, 8000 $L_{\odot}$, was used.
For simplicity,  we adopt this value for a distance of this object
in this paper.

Paper I gave two reasons for the central star to be an evolved object:
(1) the presence of  deep CO first overtone absorption bands 
at 2.3 $\mu$m as usually seen in late MIII stars, and
(2) no nearby starforming activity such as often seen in YSOs. 
Furthermore, we can add the OH maser line property of this object:  
 the dominant OH 1612 MHz line  
accompanying weak OH 1665/1667 HMz lines  (Lewis 2001; private communication),
which is often used to separate evolved objects from YSOs (Caswell 2000).
All of these facts strongly support the evolved star interpretation of the central star.

There is  evidence for O-rich material
surrounding a class of carbon stars (\cite{wil86}; \cite{nak87};
\cite{leb90}; \cite{jia00}). These materials are considered to be the 
gas and dust which were ejected in the O-rich AGB era of the central star,
but were trapped by a hypothetical binary companion (\cite{lly90}).
When the central star changes the atmospheric composition from O-rich to C-rich,
the presence of the O-rich material would observationally become 
evident due to their chemistry difference.  
In fact, a sharp narrow CO line profile ($\sim 1$ km s$^{-1}$) having slightly broader 
wing emission ($\sim 5$ km s$^{-1}$), which is similar to the narrow 
and broad components of IRAS 19312+1950, 
has been found  in the silicate carbon star, BM Gem (\cite{kah98}; \cite{jul99}).
In order to explain  O-rich materials 
surrounding silicate carbon stars,  a model of orbiting  molecular reservoirs 
has been proposed (\cite{lly90}; \cite{jul99}). 
In contrast, \citet{ker99} found
O-rich examples (semiregular and irregular variables) 
of the narrow ($\lesssim$ 5 km s$^{-1}$) feature 
superposed on the broad ($\sim 20$ km s$^{-1}$) component in CO lines.
These may be interpreted  as a result of  either multiple winds
or a Keplerian disk (\cite{ber00}).

It is possible that we are looking these trapped
materials as a narrow-component cloud [for example, \citet{kah98}].
Because the narrow and broad components found in the
 IRAS 19312+1950 line profiles have a similarity with the CO line features 
found in  these stars,  the model  of orbiting  molecular reservoirs must be
pursued first in the present case too.  However, as discussed in the following sections,  
the derived total mass of the gas ($\sim 25 \; M_{\odot}$ within 15$''$ radius), which is necessary to explain the narrow component of IRAS 19312+1950,  seems too large as an orbiting molecular reservoir. 

\begin{figure}
  \begin{center}
    \FigureFile(80mm,120mm){fig7.eps}
  \end{center}
  \caption{
  Line profiles of NH$_{3}$, C$^{34}$S, C$^{17}$O, and $^{13}$C$^{18}$O lines.
  }\label{fig7}
\end{figure}

\subsection{The broad component: molecular abundances}
In order to obtain molecular abundances, numerical calculations 
based on the large-velocity-gradient (LVG) model were performed.
Here the broad-component emission was assumed to come from the expanding
envelope of an AGB/post-AGB object. 
The numerical code developed by \citet{deg90} was used to compute line intensities
of linear molecules in the expanding shell.
The code was modified to involve 
SO molecule which has a slightly complex energy-level structure;
transition probabilities (Einstein A coefficients etc.) of SO were 
taken from \citet{tie74} [see the level diagram in \citet{omo93}].
In this numerical code, 
the kinetic temperature of the envelope is calculated
by taking into account the adiabatic and line (CO, CI, and CII) 
cooling and the gas--dust drift and CI ionization heating
(see the detail in \cite{deg90}). Dissociations of 
CO, HCN, SiO, and SO due to interstellar UV radiation 
are taken into account;  self absorptions of CO 
UV lines  are also taken into account. 
However, photodissociation rates of SiO and SO 
are not well known. Therefore, we assumed that the
photodissociation rates are the same as the
CO dissociation rate (without self-absorption).

We assumed that the envelope is expanding with a constant 
velocity of 25 km s$^{-1}$. 
The calculation starts from the inner radius of $1.5\times 10^{16}$ cm and
stops at the outer radius of $1.87 \times 10^{17}$ cm 
(corresponding to a 5$''$ radius from the central star).
The mass loss rate (to be constant), 
kinetic temperature at the inner radius, and molecular abundance
were free parameters and the best-fit model to the observational data 
was sought. The results are given in table 7, where molecular species, 
abundance of the observed molecule with the broad component,
frequencies, transitions, and the observed main-beam brightness temperature,
and the calculated main-beam brightness temperature were shown for the model 
with the mass loss rate of $2.6 \times 10^{-4} M_{\odot}$ yr$^{-1}$.
In this model, the main-beam brightness temperature was
determined mostly from  line brightness temperature at around 
$10^{17}$ cm with $N_{\rm H_{2}}=1.1\times 10^{4}$ cm$^{-3}$
(except the CO $J=1$--0 transition).
Because photodissociation of molecules occurs slightly outside 
of this radius, a slight uncertainty of the photodissociation rate
in various molecules except CO does not influence much to the results.
The obtained SO abundance, $5.5\times 10^{-7}$, is similar to those
found in normal O-rich envelopes (\cite{gui86};\cite{omo93}).

 {\setlength{\tabcolsep}{2pt}\footnotesize
 \begin{table}
  \caption{Model fit for the broad component for mass loss rate of $2.6 \times 
  10^{-4} M_{\odot}$ yr$^{-1}$}\label{tab7}
     \begin{center}
  \begin{tabular}{lrrlrr}
  \hline\hline
Mol.&Abundance& Frequency&Transition&$T^{\rm obs}_{\rm MB}$&$T^{\rm model}_{\rm MB}$\\
 & (per H$_{2})$&(GHz)& &(K)&(K)\\
	\hline
SiO&$1.1\times 10^{-7}$&86.847&2--1 $v=0$&0.18&0.18 \\
HCN&$3.9\times 10^{-7}$&88.632&1--0 &0.41&0.41\\
HCO$^{+}$&$5.7\times 10^{-8}$&89.188&1--0&0.12&0.12 \\
$^{13}$CO&$4.0\times 10^{-6}$&110.201&1--0&$<0.1$ &0.09\\
CO&$1.3\times 10^{-4}$&115.271&1--0&1.79&1.79 \\
	\hline
SO&$5.5\times 10^{-7}$&86.094&2(2)--1(1)&0.05&0.04 \\
   &                  &99.299&3(2)--2(1)&0.22&0.20 \\
   &                  &109.252&2(3)--1(2)&0.05&0.10 \\
	\hline
	 \end{tabular}
	 \end{center}
\end{table} 
}

For the case of HCO$^{+}$, the abundance of this molecule
is assumed to be constant through the envelope in the present calculation. Because
the molecular abundances of HCN and HCO$^{+}$ in the O-rich 
circumstellar envelopes were somewhat controversial
(\cite{mor87}; \cite{deg90}), we also computed the 
expected HCO$^{+}$ abundance in the same code used in \citet{deg90}.
The maximum HCO$^{+}$ abundance can reach to 
$1.5\times 10^{-5}$ per H$_{2}$ at the radius 
$2.3\times  10^{17}$ cm (if  mass loss occurs continuously beyond $1.87 \times 10^{17}$ cm). 
Because there are still some arguments
in the H$_{3}^{+}$ dissociative recombination rate (\cite{ama88}), 
this value is probably available maximum. 
The obtained HCO$^{+}$ abundance  is
less than the value obtained from the present photo-chemical model 
(see also \cite{gla96}).  Furthermore, clumpiness in the outflowing envelope
strongly influences the photo-chemical products (\cite{red03}).
The spherical model with outer radius of
$1.87\times 10^{17}$ cm may be inadequate for explaining HCO$^+$.
 In fact, a bipolar structure has been found 
 for the HCO$^+$ emission distribution (\cite{nak04b}). 
 
\subsection{The narrow-component cloud: molecular abundance}

{\setlength{\tabcolsep}{2pt}\footnotesize
\begin{table*}
  \caption{Molecular abundance for narrow component}\label{tab:table8}
   \begin{center}
  \begin{tabular}{llcrrr}
  \hline\hline
Molecule & Abundance &   Freq. & Transition & $T_{MB}^{\rm obs}$ & $T_{MB}^{\rm calc}$ \\
               &  (per H$_2$) &  (GHz)   &  (K)                         &  (K)  \\
	\hline
CO    &  $1.3 \times 10^{-4}$  & 115.271 &  1--0 &   6.83  & 6.83 \\
HCN &  $4.0 \times 10^{-9}$   &    88.631 &1--0  &    0.77  & 0.77 \\
HCO$^+$ &  $1.2 \times 10^{-9}$   &   89.189 & 1--0  &    0.53  & 0.53 \\
HNC &  $3.4 \times 10^{-9}$   &   90.663 &1--0  &     1.12  & 1.12 \\
HC$_3$N&  $7.0 \times 10^{-9}$   &  100.076 &11--10  &    0.31  & 0.31 \\
                &                                        &     109.174  & 12--11 & 0.19  & 0.22 \\
CS &  $1.0 \times 10^{-8}$   &    48.991 & 1--0  &  0.85  & 1.23 \\
     &                                      & 97.981   &   2--1   &    1.77  & 1.66 \\
SO &  $5.5 \times 10^{-9}$  &   86.094   &  2(2)--1(1) &    0.05  & 0.27 \\
       &                                    &    99.300  & 3(2)--2(1)    & 0.77  & 0.77 \\ 
      &                                &   109.252    &  2(3)--1(2)    &  0.17  & 0.42 \\
   \hline \\
  \end{tabular}
\end{center}
\end{table*} }

In order to obtain  molecular abundances in the narrow component cloud,
we  performed numerical calculations under the LVG model. Though
the LVG model was a choice for a matter of convenience here, more sophisticated,
but formidable radiative-transfer model involving clumpiness and turbulence 
with velocity gradient (\cite{par95}) must be pursued for the narrow component cloud as a next step;  
there is evidence that the spectral line of the narrow 
component is broadened  by turbulence as mentioned in section 2.3. However, the present data,
which was  obtained in a single-dish observation,  do not provide much information on spatial 
distribution of molecular emission. Then, it seems meaningless 
to argue on differences of outcomes between the LVG and turbulence models. 
Therefore, we used the LVG model for the narrow 
component cloud with inner and outer radii,  $1.87 \times 10^{17}$ cm and
$5.6  \times 10^{17}$ cm  (corresponding to 5 and 15$''$ radii), respectively. The uniform
expansion rate, $dV/dr=3.2\times 10^{-13}$ s$^{-1}$,  was adopted
giving an expansion velocity of 2 km s$^{-1}$ at the outer boundary.
Radiative interaction with the central expanding core (broad component) was also
taken into account as if the core emits continuum emission, but it turned out to be negligible.
The number density was assumed to decrease with radius as $r^{-3}$, and
the absolute value of the density and the molecular abundances were adjusted 
to fit calculated to observed line intensities.  
For simplicity, the calculations were restricted for linear-type molecules (except SO).
We solved  level populations up to $J=6$ for CO, HCN, HCO$^+$, and HNC,
to $J=8$ for CS , and to $J=15$ for HC$_3$N at different radii and calculated the
brightness temperatures. Multiplying the telescope beam pattern, 
we obtained  the molecular abundances to fit the calculated to the  main-beam brightness 
temperatures of  observed  lines.  Kinetic temperature in the model (uniform) was determined 
to be 13 K by adjusting the calculated $^{12}$CO line intensity to the observed intensity.  
The results are shown in  table 8. It suggests that the abundances of
HCN and SO are two orders of magnitude lower than those for the broad component
(table 7), making the discrete difference of the molecular composition
at the boundary between narrow and broad components. The total mass of the narrow component cloud 
in this model is 24.4 $M_{\odot}$ (within 15$''$ radius), which is consistent 
with the previous estimation
of the mass of this dust cloud based on a $^{18}$CO column density (\cite{nak04}).

\subsection{isotopic abundance}
Another important clue to guess the origin of cool material surrounding the object
is  isotopic abundance. For this purpose, we obtained the isotope abundance ratios
for the narrow-component cloud
with the same LVG model used in the previous section.
Table 9 summarizes the isotope line-intensity (peak-intensity) ratios and 
calculated abundance ratios (in the 2nd and 5th columns, respectively).  For comparison,
terrestrial and interstellar isotope 
abundance ratios are given in the last two columns.  
If  lines are optically thin, the line intensity ratio approximately 
gives the isotope abundance ratio. However, because strong lines,  
as $^{13}$CO $J=1$--0, HCO$^{+}$ $J=1$--0 and CS
$J=2$--1, tend to be saturated in intensity (optically thick),
they only give a lower limit of abundance ratio.

The ratio,  C$^{18}$O/$^{13}$C$^{18}$O, is very often
regarded to give the $^{12}$C/$^{13}$C abundance ratio  in dark clouds 
in a good approximation [for example, see \citet{lan93}] 
because these lines are optically thin.
 In fact, in the LVG model computations, the line intensities
 of the minor CO isotopic species vary almost proportional to the abundance  of
 CO [times $(dV/dr)^{-1}$ too] [e.g.,  figure 17 of \citet{cas90}],
as far as it is optically thin. 
We obtained the ratio $^{12}$C/$^{13}$C$\sim 27\pm 9$
from the C$^{18}$O/$^{13}$C$^{18}$O line ratio in table 9.
It is well known that $^{13}$C is enriched in envelopes of some carbon stars 
[for example, see \citet{woo03}].  Furthermore,  \citet{jul88} found low line intensity ratio 
of $^{12}$CO to $^{13}$CO $J=$1--0 transition
in several optical carbon stars which exhibit $^{13}$C-richness.
The observed abundance ratios are approximately between 3 and 80
for evolved objects [for example, \citet{sch00}].  No enrichment of $^{13}$C
compared with the case of interstellar clouds (in table 9) indicates 
that the material is unlikely to be an ejecta from a (supposedly) dead,  C-rich AGB companion.

However,  we have to remind that 
the detection of $^{13}$C$^{18}$O $J=1$--0 line was marginal
(see the lowest panel of figure 7). Taking into account the uncertainty of 
the temperature calibration and  pointing errors of the 45-m telescope at 110 GHz
due to wind etc., we think that the observed value involves 
uncertainty of a factor of about 2.  Furthermore, if we consider the model dependency
of the CO line intensity, especially in turbulent clumpy clouds, and of emission size difference
in  C$^{18}$O and $^{13}$C$^{18}$O  lines,  we think 
that the conclusion is somewhat provisional.
 
\citet{kah92} observed several carbon stars and planetary nebulae, 
and found that the $^{18}$O/$^{17}$O ratio is less than unity.
However, the present case is apparently different from such C-rich objects;
the C$^{18}$O/C$^{17}$O $\sim 5$ in table 9.  Concerning on
$^{17}$O, we can safely conclude that it is not appreciably enriched 
in the narrow-component cloud as in these C-rich objects.
Rather,  the $^{18}$O/$^{17}$O ratio coincides with the interstellar value.
In conclusion, we found no evidence of the isotope enrichment/deficiency compared 
with the interstellar values in the present observation. 

{\setlength{\tabcolsep}{2pt}\footnotesize
\begin{table*}
  \caption{Isotope line-intensity and abundance ratio (narrow component)}\label{tab:table9}
  \begin{tabular}{lcccccc}
  \hline\hline
Molecule & Peak I. ratio & Uncertainty & Transition & Calc. abu. ratio  & Terrestrial ratio$^1$ & Interstellar ratio\\
	\hline
$^{13}$CO/$^{13}$C$^{18}$O  &  66  & $\pm$ 20  & 1--0 & 498  &   489   &  489$^2$   \\
$^{13}$CO/C$^{18}$O                &   3.6  & $\pm$ 0.3  & 1--0  & 19  &    5.43  &   7--25$^3$  \\
 C$^{18}$O/$^{13}$C$^{18}$O  & 18  & $\pm$  6  & 1--0  & 27 &     89.1     &  20--70$^4$\\
C$^{18}$O/C$^{17}$O                &  4.1   & $\pm$ 0.7  & 1--0 & 5.7  &     5.5  &     3.2$^2$     \\
 HCO$^+$/H$^{13}$CO$^+$  &  2.2 &  $\pm$ 0.5  & 1--0 & 10  &     89.1 &  $\sim 50^5$ \\
 CS/C$^{34}$S                         &   4.8  & $\pm$ 1.0   & 2--1  & 9.1   &    22.5 &    24$^5$ \\
   \hline \\
  \end{tabular}
  \\
$^1$ from appendix VII of \citet{tow75}. \\
$^2$ \citet{wil92} \\
$^3$ \citet{gre94} \\
$^4$ \citet{lan93}, \cite{sav02}; see references therein. \\
$^5$ \citet{luc98} \\
$^6$ \citet{chi96} \\
\end{table*} }

\subsection{Narrow component: what is it ? }

The detections of SO, HCN,  HCO$^+$, and NH$_3$ first
suggested a resemblance of chemistry between IRAS 19312+1950
and OH231.8+4.2 (a proto-planetary nebula), leading a conclustion
that IRAS 19312+1950 was a protoplanetary nebula. 
In fact, the first three molecules were observed to
have a broad component as well.
However, we recognized several differences 
between the molecular species found in IRAS 19312+1950 and OH231.8+4.2.
Firstly, SO$_{2}$ molecules are quite deficient in IRAS 19312+1950
(see table 4), but were abundant in OH231.8+4.2 (\cite{mor87}), although
SO lines are detected in both. The line intensity ratio of the
SO$_{2}$ 3(13)--2(02) (104.029 GHz) to SO  2(2)--1(1)
line (86.094 GHz) is 6.2 in OH231.8+4.2 (\cite{mor87});
the abundance ratio of SO$_2$ to SO  is about 4--7 (\cite{cla00}), and
somewhat similar values have also been found in O-rich AGB stars 
(\cite{sah92})  (except 2 supergiants, NML Cyg and VY CMa, which seem to be SO$_2$-poor
relative to SO; \cite{omo93}). 
However,  this intensity ratio is $\lesssim 0.2$ in
IRAS 19312+1950 (in the condition that the marginally detected SO$_2$ line 
is regarded to be a narrow component). 
The deficiency of SO$_2$ relative to SO in IRAS 19312+1950 could be attributed 
to a difference in O/C abundance ratio  between two stars, or
to less developement of shock 
compared with OH231.8+4.2 (e.g., \cite{cla00}).
Secondly, the NH$_{3}$ (1,1) and (2,2) inversion lines were detected
in IRAS 19312+1950. The line intensity ratio
suggests the rotational temperature of the concerning levels
 (both para-NH$_{3}$)  of 19 K for the narrow component cloud.
 For the case of  OH 231.8+4.2,  the line intensities suggested
 the NH$_{3}$ rotational  temperature of $(27\pm 4)$ K (\cite{mor87}),
 indicating that the narrow-component cloud is slightly cooler than 
 the envelope of OH231.8+4.2
  
Rich molecular lines detected toward IRAS 19312+1950
 (as a narrow component) rather resemble to the lines usually found 
in dark clouds. For example, in L134N, the SO lines have been 
found to be strong but not much SO$_{2}$ (\cite{swa89}; \cite{dic00}).
Therefore, the deficiency of SO$_2$ to SO can be explained naturally
if we consider that the narrow-component cloud is a dark cloud as  L134N.

The detection of  CH$_3$OH  in this paper also favors 
 the dark-cloud interpretation  of
the narrow component cloud. The detected  CH$_3$OH  lines at 96.74 GHz
are thermal emission and have been found widely in dark (\cite{fri88})
and translucent clouds (\cite{tur98}). Note that 
a search for the 6.7 GHz CH$_3$OH $5_1$--$6_0$ A$^+$ line, one of the strongest
CH$_3$OH maser lines,  has been negative in IRAS 19312+1950 
(\cite{mac98}; \cite{szy00}), although numerous maser lines
of  CH$_3$OH have been detected normally in dense massive star forming clouds
(in this sense, the cloud toward IRAS 19312+1950 is not a massive star forming region).
Though CH$_3$OH  could be produced in circumstellar envelopes of 
 O-rich stars (\cite{cha95}),   it has never been detected 
 in  evolved stars to the authors' knowledge [see \citet{lat96} ; \citet{cha97}]. 

Time-dependent chemical models of low-mass dark-cloud cores
(\cite{nej99}) gave a result that abundances of C-bearing complex molecules
as  CH$_3$OH and CH$_3$CN  are peaked at a few times $10^5$ years,
while the peak abundance of SO and SO$_2$  are attained at later times of 
$\sim 10^6$ years (though these time scales are dependent  on the density of the cloud).   
Note that the chemistry of CH$_3$OH and CH$_3$CN in translucent clouds  is 
still quite controversial [see an excellent summary in \citet{tur98} 
on gas-phase and grain-surface chemistry], 
and the confinement of their chemical evolution at relatively earlier time 
in the \citet{nej99}'s model was critically affected 
by cosmic-ray-induced photoionization and photodissociation processes.

It is useful to compare the time scale of molecule formation with the other 
physical time scales of the envelope. The time scale
of expanding gas with 10 km s$^{-1}$ crossing the
envelope (15$'' \sim 5.8\times 10^{17}$ cm)
is as short as $1.8\times 10^4$ y.
Therefore, even if we interpret the narrow-component 
cloud being the stellar ejecta, we need relatively long-lived ($\sim 10^6$ years)
trapped materials to create CH$_3$OH. The period of circulation around a 5 $M_{\odot}$
star at  10$''$ radius is $1.8\times 10^6$ y. Therefore,
it is possible to create the complex molecules as CH$_3$OH
within one orbital motion of ejected gas. Because the life time of the thermal-pulse AGB era
is $\lesssim 10^6$ y for a progenitor mass of $>4 \; M_{\odot}$ (\cite{blo95}),
the ejecta must come from the companion, not from the
AGB/post-AGB star which is seen now as a central star
of the mass of about 5 $M_{\odot}$ (Paper I). 

It is possible to consider that material of the narrow-component cloud
is a remnant of the dark cloud which encountered
with this object several million years ago.
Such encounter with a dark cloud is considered to occur
relatively frequently in the solar neighborhood  
(in  every $\sim$10 Myr per a star; \cite{gri97}; \cite{fre03}).
Taking into account the chemical-evolution time scale of
dark clouds, we estimate that such encounter must have 
occurred in the last $10^6$ yr ago for IRAS 19312+1950. 
In this case, the encountered
cloud cannot be very far  away from the object at present; for example,
it must be located within 5$'$ from IRAS 19312+1950,
provided that the relative velocity between the star and 
the cloud is 10 km s$^{-1}$.  
This is also an  attractive hypothesis
in view of an explanation for silicate carbon stars and narrow
CO line profiles of short-period variable stars.
However, the large mass of the narrow-component material 
(as much as $\sim 20 M_{\odot}$) again causes a problem 
in this interpretation;  It must be quite difficult to
accrete the mass more than the star mass during the passage of 
such dark cloud and to keep it around the object during a few Myr.

Note that  brightness of the central star  at near-infrared bands
($K=6.3$ in IRAS 19312+1950 ; Paper I) is so different from  class-0 YSOs  
as  L134N or L1551 IRS5. 
These objects  usually do not have any bright near-infrared counterparts.
For example, L1551 IRS5 does show very faint patchy structure
in the K-band images [\cite{hod94}; or see \citet{mom98} for the overlay image with the 
C$^{18}$O $J=1$--0 map]. As far as we see 2MASS images
of the similar type of objects as L134N (\cite{dic00}) and  IRAS 16293$-$2422 (\cite{bla94}),
we find no near-infrared counterpart. Though
HH objects as B5 IRS 1 (\cite{ful91}; K=11.21) tend to have
brighter near-infrared counter parts (\cite{rei01}), these objects have
well developed CO outflow too. On the other hand, the size of high-velocity flow of 
IRAS 19312+1950 (if the broad component is interpreted as a bipolar flow 
observed in YSOs) is  small compared with the size of
narrow component cloud ($\sim 20''$),  suggesting again a difficulty of such
interpretation that the central star is a young stellar object.

From above arguments, we are forced to conclude  
that the narrow-component cloud is a small dark cloudlet.
However, because the presence of a ring structure and 
spurs seen in figure 1 indicates a physical relation 
between the central star and the surrounding cloud,
the narrow component must be a part of the object, 
which is physically associated with the central star,
even though they may not be the gas ejecta from 
the central star or hypothetical companion. 
In order to check the possibility of YSO further,  it might be nice
to observe  CO and CO$_2$ ice absorption bands  at 4.2--4.7  $\mu$m
(as a revelation of molecular clouds: \cite{gib00}) 
 and crystalline  H$_2$O  at 44/62 $\mu$m 
(as a revelation of post-AGB outflows; \cite{mal03})
for the future work .

It is probably the best to perform mm-wave interferometric
observations with high spatial resolution in order to clarify
the physical relation between the narrow and broad components.
Preliminary results of interferometric mapping of the CO  $J=1$--0/2--1
emissions with Nobeyama and Berkeley-Illinois-Maryland Arrays gave that 
the broad component was partially resolved in $\sim 3 ''$ beam,
but the narrow component is extended to about more than 10$''$ in size.
These results are consistent with the 45-m mapping observations (Paper I),
 and they fit well to the present interpretation that
 the broad component is the expanding envelope of the AGB/post-AGB object
which is surrounded with a cool cloud at the distance of 2.5 kpc.

\section{Conclusion}

We have detected a variety of molecular lines of
O-, C- and N-bearing molecules toward IRAS 19312+1950,
a bipolar nebula. The line profiles consist of
narrow and broad components of the widths of $\sim$4 and $\sim$30 km
s$^{-1}$, respectively. From the parabolic line shape,
the broad component is inferred to 
originate from the circumstellar envelope of this object.
The LVG calculations suggest the mass loss rate of about 
$2.6 \times 10^{-4} M_{\odot}$ yr$^{-1}$ for this object.
However, the origin of the narrow component is still a puzzle;
from the low temperature and variety of molecules detected,
it is suggested to be a dark cloudlet with a mass of about 20 $M_{\odot}$. 
If it were a trapped material by a hypothetical companion,
the H$_2$/CO mass ratio must be a factor of 10 less than
the value normally assumed.  If this is really a dark cloud with a mass
more than 20 $M_{\odot}$, how the AGB/post-AGB star can accommodate
a  cloud more massive than the star mass ? Furthermore,
if this is a low-mass young stellar object with SiO masers more than a few kpc away,
it is hard to explain the apparent size of the bright nebulosity
and luminosity of the central star.

It is hard to provide a reasonable interpretation for this object
except considering somewhat exotic, previously unknown, star model
as a hypergiant (\cite{jur01}), or possibly an AGB/post-AGB object 
with the huge orbiting gas reservoir in which the H$_2$/CO ratio is small compared 
with a normal ratio. Another extreme is a low-mass young stellar object with SiO masers
with yet undeveloped CO outflow but shining the surrounding dust
of the size of 30$''$.  Neither of them could be acceptable, except with excuse 
if they were extreme objects.

\vspace{5mm}

The authors thank Dr. M. Tamura, and K. Murakawa for providing them with the SIRIUS image
of IRAS 19312+1950. They also thank Dr. T. Umemoto for useful suggestions, and 
students (H. Fukushi, and Y. Tamura) for help of observations.


\end{document}